\ifdefined\manuscriptlayout\else\def\manuscriptlayout{preprint}\fi
\expandafter\documentclass\expandafter[\manuscriptlayout,aip,pop,floatfix,amsmath,amssymb]{revtex4-2}
\usepackage{graphicx,booktabs,amsthm}
\makeatletter
\def\@hwromreprint{reprint}%
\ifx\manuscriptlayout\@hwromreprint
  \let\hwromtwocolumn\@firstoftwo
\else
  \let\hwromtwocolumn\@secondoftwo
\fi
\newsavebox{\hwrommathbox}
\newcommand{\hwrombox}[1]{%
  \hwromtwocolumn{%
    \begingroup
    \sbox{\hwrommathbox}{$\displaystyle\boxed{#1}$}%
    \ifdim\wd\hwrommathbox>\dimexpr\linewidth-3em\relax
      \mathord{\resizebox{\dimexpr\linewidth-3em\relax}{!}{\usebox{\hwrommathbox}}}%
    \else
      \mathord{\usebox{\hwrommathbox}}%
    \fi
    \endgroup
  }{\boxed{#1}}%
}
\hwromtwocolumn{%
  \setcounter{dbltopnumber}{3}%
  \setlength{\@dblfptop}{0pt}%
  \setlength{\@dblfpsep}{12pt plus 2pt minus 2pt}%
  \setlength{\@dblfpbot}{0pt plus 1fil}%
}{%
  \addtolength{\textheight}{3pt}%
}
\makeatother
\usepackage[colorlinks=true,linkcolor=blue,citecolor=blue,urlcolor=blue]{hyperref}
\graphicspath{{figures/}}
\theoremstyle{definition}

\DeclareFontShape{OT1}{cmss}{bx}{it}{<->ssub * cmss/bx/n}{}
\begin{document}
\title{Recovering hysteresis in reduced models of plasma turbulence: a Mori--Zwanzig-inspired closure}
\author{Ameir Shaa}
\email{ameirshaa.akberali@ntu.edu.sg}
\affiliation{School of Physical and Mathematical Sciences, Nanyang Technological University, Singapore 637371, Singapore}
\author{Kyungtak Lim}
\affiliation{School of Physical and Mathematical Sciences, Nanyang Technological University, Singapore 637371, Singapore}
\author{Claude Guet}
\affiliation{School of Physical and Mathematical Sciences, Nanyang Technological University, Singapore 637371, Singapore}
\author{Xavier Garbet}
\affiliation{School of Physical and Mathematical Sciences, Nanyang Technological University, Singapore 637371, Singapore}
\affiliation{CEA, IRFM, F-13108 Saint-Paul-lez-Durance, France}
\date{\today}
\begin{abstract}
Hysteretic transitions between turbulent and strongly zonal-flow-dominated states provide an important test for reduced models of plasma turbulence and are closely connected to L–H transition physics. The 12-mode truncation of the modified Hasegawa–Wakatani system developed by Guillon and Gürcan retains this transition but underestimates the hysteresis observed in direct numerical simulations. Guided by the Mori--Zwanzig framework, we construct a phenomenological closure for the influence of discarded modes. We coarse-grain the discarded spectrum into a reservoir, then model the feedback of the discarded spectrum on the retained modes. Without increasing the number of retained modes, the closure recovers
nearly all of the hysteresis width missing from the unclosed truncation
and places both transition thresholds close to the DNS values. As a test of whether the reduced-order model (ROM) generalizes beyond hysteresis, we additionally examine particle flux. The ROM substantially improves on the 12-mode truncation, agreeing with DNS at low adiabaticity but underestimating transport at high adiabaticity. These results show that a low-dimensional representation of discarded-mode feedback can improve predictions of hysteretic turbulence–flow transitions.
\end{abstract}
\maketitle
\raggedbottom
\section{Introduction}\label{sec:intro}

Transitions between turbulent and strongly sheared-flow states are central to confinement regulation in magnetized plasmas. Drift-wave (DW) turbulence generates zonal flows (ZFs) through nonlinear energy transfer, while the resulting shear suppresses turbulence and regulates cross-field transport~\cite{horton1999,biglari1990,diamond2005,lin1998,fujisawa2009}. This turbulence--flow feedback can support distinct dynamical states and hysteretic transitions, contributing to the physics of the low- to high-confinement (L--H) transition~\cite{diamond1994,kim2003,schmitz2012,schmitz2017,miki2012,malkov2009}. In particular, experiments have identified turbulence-driven ZFs and flow shear as important components of the transition to H-mode confinement~\cite{schmitz2012}.

The modified Hasegawa--Wakatani (MHW) system provides a minimal setting for studying this nonlinear interaction~\cite{hasegawa1983,wakatani1984,numata2007}. The MHW system exhibits a transition between turbulent and ZF-dominated states~\cite{numata2007}, with different transition thresholds as the adiabaticity parameter $C$ is increased and decreased~\cite{guillon2025}. Related hysteretic behavior has been observed in a warm-ion gyrofluid extension~\cite{grander2024}. Although the MHW system does not describe the full L--H transition, it allows the associated turbulence–flow feedback and hysteresis to be studied in a simplified setting.

Reduced-order models (ROMs) seek to retain these transition dynamics with fewer degrees of freedom~\cite{benner2015}. Spectral truncation may retain the modes needed for a transition while removing interactions needed to reproduce its hysteresis. For the MHW system, Guillon and G\"urcan identified a 12-mode Fourier truncation that preserves the transition between turbulent and ZF-dominated states~\cite{guillon2025}. In our comparison with \(512\times512\) direct numerical simulations (DNS), however, this unclosed truncation recovers only about \(16\%\) of the hysteresis width. Richer spectral truncations provide one route to improved fidelity~\cite{guillon_aniso2026,guillon_flux2025}, while closure and data-driven approaches represent the effects of discarded scales~\cite{li2026}. Here we ask whether a low-dimensional closure for the discarded-mode feedback can recover the missing hysteresis without increasing the number of retained Fourier modes.

The Mori--Zwanzig (MZ) formalism provides a framework for expressing the influence of eliminated degrees of freedom on the retained dynamics~\cite{nakajima1958, zwanzig1961, mori1965}. Eliminating these degrees of freedom yields exact equations containing memory and fluctuating contributions, providing a systematic starting point for reduced-order closure~\cite{chorin2000,chorin_physd2002,givon2004}. This framework has motivated closures for nonlinear and turbulent systems~\cite{stinis2015,parish2017prf,parish2017jcp,lin2021}, including plasma turbulence, where Maeyama and Watanabe showed that eliminated scales can damp, phase-shift, and excite retained fluctuations~\cite{maeyama2020}. Low-dimensional phenomenological models have shown that essential turbulence--flow transition dynamics can survive substantial reduction. Malkov and Diamond used three coupled variables representing DW turbulence, zonal flow, and the driving pressure gradient to reproduce oscillatory intermediate states, bistability, and hysteresis, including an asymmetry between the forward and reverse transitions~\cite{malkov2009}. Their approach illustrates how effective feedback laws can capture transition dynamics without resolving the underlying turbulent spectrum.

Guided by the Mori--Zwanzig framework, we construct a closure for the feedback of the discarded spectrum while preserving the original 12-mode triadic structure. We first coarse-grain the discarded spectrum into a reservoir and determine its dynamics. We then model how this coarse-grained description couples back to the retained modes. We test whether this recovers the hysteresis lost under spectral truncation and whether the resulting reduced-order model generalizes to particle flux. These comparisons assess how much of the discarded-spectrum influence can be represented without adding retained Fourier modes.

The remainder of the paper is organized as follows. Section~\ref{sec:background} introduces the MHW system and the 12-mode reduction. Section~\ref{sec:methodology} derives the resolved--unresolved energy balance and develops the closure. Section~\ref{sec:results} compares the closed model with full-field DNS. Section~\ref{sec:discussion} discusses the physical interpretation and limitations of the model. Section~\ref{sec:conclusion} summarizes the main conclusions.
\section{Modified Hasegawa--Wakatani model and 12-mode reduction}
\label{sec:background}

The MHW system provides the DNS reference, while the canonical 12-mode truncation of Guillon and G\"urcan~\cite{guillon2025} defines the resolved state. We introduce the spectral equations and transformed components needed to construct the closure, then define the transition diagnostic.

\subsection{MHW dynamics in Fourier space}

The MHW system describes resistive DW turbulence in a plane perpendicular to a uniform magnetic field $\mathbf B=B\hat{\mathbf z}$~\cite{hasegawa1983,wakatani1984,numata2007}. The coordinates $x$ and $y$ denote the radial and poloidal directions. The dynamical fields are the normalized electrostatic potential $\phi$ and density perturbation $n$ about a prescribed background density $n_0(x)$. The vorticity is $\Omega=\nabla^2\phi$, and the background density gradient $\kappa=-\partial_x\ln n_0$ supplies the free energy for the DW instability.

We decompose each field as $A=\bar A+\widetilde A$, where $\bar A(x,t)=L_y^{-1}\int_0^{L_y}A(x,y,t)\,dy$ is the zonal average. The modified system restricts the adiabatic coupling to non-zonal fluctuations~\cite{numata2007}. Here dissipation is also restricted to these fluctuations, giving
\begin{subequations}
\label{eq:mhw}
\begin{align}
\partial_t\Omega+[\phi,\Omega]
&=C(\widetilde\phi-\widetilde n)+\nu\nabla^2\widetilde\Omega,\\
\partial_t n+[\phi,n]+\kappa\partial_y\phi
&=C(\widetilde\phi-\widetilde n)+\nu\nabla^2\widetilde n.
\end{align}
\end{subequations}
Here $[\phi,f]=\partial_x\phi\,\partial_y f-\partial_y\phi\,\partial_x f$ represents $E\times B$ advection. The adiabaticity parameter $C$ controls the coupling between density and potential fluctuations, and the same dissipation coefficient $\nu$ is used in both equations.

For periodic fields, we use $f(\mathbf x,t)=\sum_{\mathbf k}f_{\mathbf k}(t)e^{i\mathbf k\cdot\mathbf x}$, where $\Omega_{\mathbf k}=-k^2\phi_{\mathbf k}$ with $\mathbf k=(k_x,k_y)$ and $k^2=k_x^2+k_y^2$. All spectral sums exclude the spatially uniform mode and include conjugate partners satisfying $f_{-\mathbf k}=f_{\mathbf k}^{*}$. In Fourier space, the non-zonal projection takes the form $\tilde A_{\mathbf k}=P_{\mathbf k}A_{\mathbf k}$, where $P_{\mathbf k}=1-\delta_{k_y,0}$ removes the zonal modes ($k_y=0$) and leaves the non-zonal modes unchanged. Writing $\mathbf u_{\mathbf k}=(\Omega_{\mathbf k},n_{\mathbf k})^\top$, the spectral equations take the form
\begin{equation}
    \frac{d\mathbf u_{\mathbf k}}{dt}=-\mathbf L_{\mathbf k}\mathbf u_{\mathbf k}+\mathbf N_{\mathbf k},
\label{eq:spectral}
\end{equation}
where $\mathbf L_{\mathbf k}$ contains the adiabatic coupling, dissipation and background-gradient drive,
\begin{equation}
\mathbf L_{\mathbf k}=
\begin{pmatrix}
\dfrac{C_{\mathbf k}}{k^2}+d_k & C_{\mathbf k}\\[6pt]
\dfrac{C_{\mathbf k}}{k^2}-\dfrac{i\kappa k_y}{k^2} & C_{\mathbf k}+d_k
\end{pmatrix},
\label{eq:Lk_full}
\end{equation}
with $C_{\mathbf k}=CP_{\mathbf k}$ and $d_k=\nu k^2P_{\mathbf k}$.

The nonlinear term describes quadratic advection,
\begin{equation}
\mathbf N_{\mathbf k}
=
\sum_{\mathbf p+\mathbf q=\mathbf k}
(\mathbf p\times\mathbf q)\phi_{\mathbf p}
\begin{pmatrix}
\Omega_{\mathbf q}\\
n_{\mathbf q}
\end{pmatrix},
\label{eq:Nk}
\end{equation}
where $\mathbf p\times\mathbf q=p_xq_y-p_yq_x$ is the geometric coupling factor. For zonal modes, $k_y=0$ and $P_{\mathbf k}=0$, giving $\mathbf L_{(k_x,0)}=\mathbf0$. Under the chosen dissipation prescription, these modes therefore evolve solely through nonlinear coupling. Two zonal inputs have $\mathbf p\times\mathbf q=0$ and do not interact directly. Every nonzero advective contribution thus requires at least one non-zonal input, referred to here as a DW mode.

\subsection{Adiabatic and nonadiabatic components}
\label{sec:branch_variables}

Diagonalizing the coupling--dissipation part of $\mathbf L_{\mathbf k}$ (Eq.~\eqref{eq:Lk_full} with $\kappa=0$) separates each non-zonal fluctuation into a nonadiabatic component relaxed by the adiabatic coupling and an adiabatic component on which that coupling vanishes. Its eigenvectors are $\mathbf v_+=(1,1)^\top$ and $\mathbf v_-=(1,-1/k^2)^\top$, with eigenvalues $\lambda_+=C_{\mathbf k}(1+k^2)/k^2+d_k$ and $\lambda_-=d_k$. Writing $\mathbf V=(\mathbf v_+,\mathbf v_-)$, the transformed state $\widetilde{\mathbf u}_{\mathbf k}=\mathbf V^{-1}\mathbf u_{\mathbf k}$ has components
\begin{subequations}
    \label{eq:branch_amplitudes}
    \begin{align}
    \widetilde u_{+,\mathbf k}
    &=\frac{\Omega_{\mathbf k}+k^2n_{\mathbf k}}{1+k^2},\\
    \widetilde u_{-,\mathbf k}
    &=\frac{k^2(\Omega_{\mathbf k}-n_{\mathbf k})}{1+k^2}.
    \end{align}
\end{subequations}
The $(+)$ amplitude is proportional to $n_{\mathbf k}-\phi_{\mathbf k}$ and therefore measures the departure from the adiabatic density response. The $(-)$ amplitude is proportional to $\Omega_{\mathbf k}-n_{\mathbf k}$, the generalized-vorticity. It is unaffected by the adiabatic coupling, damped only by $d_k$. The same transformation is used for zonal modes, although their adiabatic coupling vanishes and neither component is linearly relaxed.

Restoring the background gives the full linear operator
\begin{equation}
\begin{gathered}
\mathbf M_{\mathbf k}=\mathbf V^{-1}\mathbf L_{\mathbf k}\mathbf V
=\begin{pmatrix}
\lambda_+-i\omega_{\mathbf k}&-i\omega_{\mathbf k}\\
i\omega_{\mathbf k}&\lambda_-+i\omega_{\mathbf k}
\end{pmatrix},\\
\omega_{\mathbf k}=\frac{\kappa k_y}{1+k^2}.
\end{gathered}
\label{eq:Mk}
\end{equation}
The background-gradient drive mixes the two components for $k_y\neq0$, allowing DW growth despite the non-negative coupling--dissipation rates $\lambda_\pm$. For zonal modes, $\mathbf M_{(k_x,0)}=\mathbf0$, consistent with their evolution through nonlinear interactions~\cite{numata2007,gurcan2015}.

The transformed spectral system is
\begin{equation}
\frac{d\widetilde{\mathbf u}_{\mathbf k}}{dt}
+\mathbf M_{\mathbf k}\widetilde{\mathbf u}_{\mathbf k}
=\mathbf c_{\mathbf k},
\qquad \mathbf c_{\mathbf k}=\mathbf V^{-1}\mathbf N_{\mathbf k}.
\label{eq:mhw_spectral_eigenbasis}
\end{equation}
Equation~\eqref{eq:mhw_spectral_eigenbasis} provides the starting point for the resolved--unresolved energy balance and closure in Section~\ref{sec:methodology}.

\subsection{Canonical 12-mode reduction}

We adopt the canonical 12-mode Fourier truncation of Guillon and G\"urcan~\cite{guillon2025}. The retained modes define the resolved state, and the remaining modes form the unresolved spectrum. The retained wavevectors are specified by

\begin{equation}
\begin{split}
\mathcal R_+=\{&(1,0),(2,0),(0,1),(0,2),\\
&(\pm1,1),(\pm1,2),(\pm2,1),(\pm2,2)\},
\end{split}
\label{eq:Rplus}
\end{equation}
where $(m_x,m_y)$ denotes $\mathbf k=(m_xq,m_yk_{y0}/2)$. Here $k_{y0}=k_{y0}(0.1)$ is the reference poloidal wavenumber at $C=0.1$, and $q=0.6k_{y0}$. This lattice remains fixed throughout the $C$ sweep, whereas the injection wavenumber entering the DNS dissipation prescription in Eq.~\eqref{eq:nu_dns} varies with $C$.

The set $\mathcal R_+$ contains twelve independent wavevectors, comprising two zonal and ten DW modes. Each wavevector carries both vorticity and density amplitudes. Including conjugate partners gives
\[
\mathcal R=\mathcal R_+\cup(-\mathcal R_+),
\qquad
\mathcal U=\mathcal K\setminus\mathcal R,
\]
where $\mathcal K$ is the complete nonzero Fourier lattice associated with the retained wavevector spacings. The reality condition determines the amplitudes at $-\mathbf k$ from those at $\mathbf k$.

\begin{figure*}[htbp]
\centering
\includegraphics[width=0.6\textwidth]{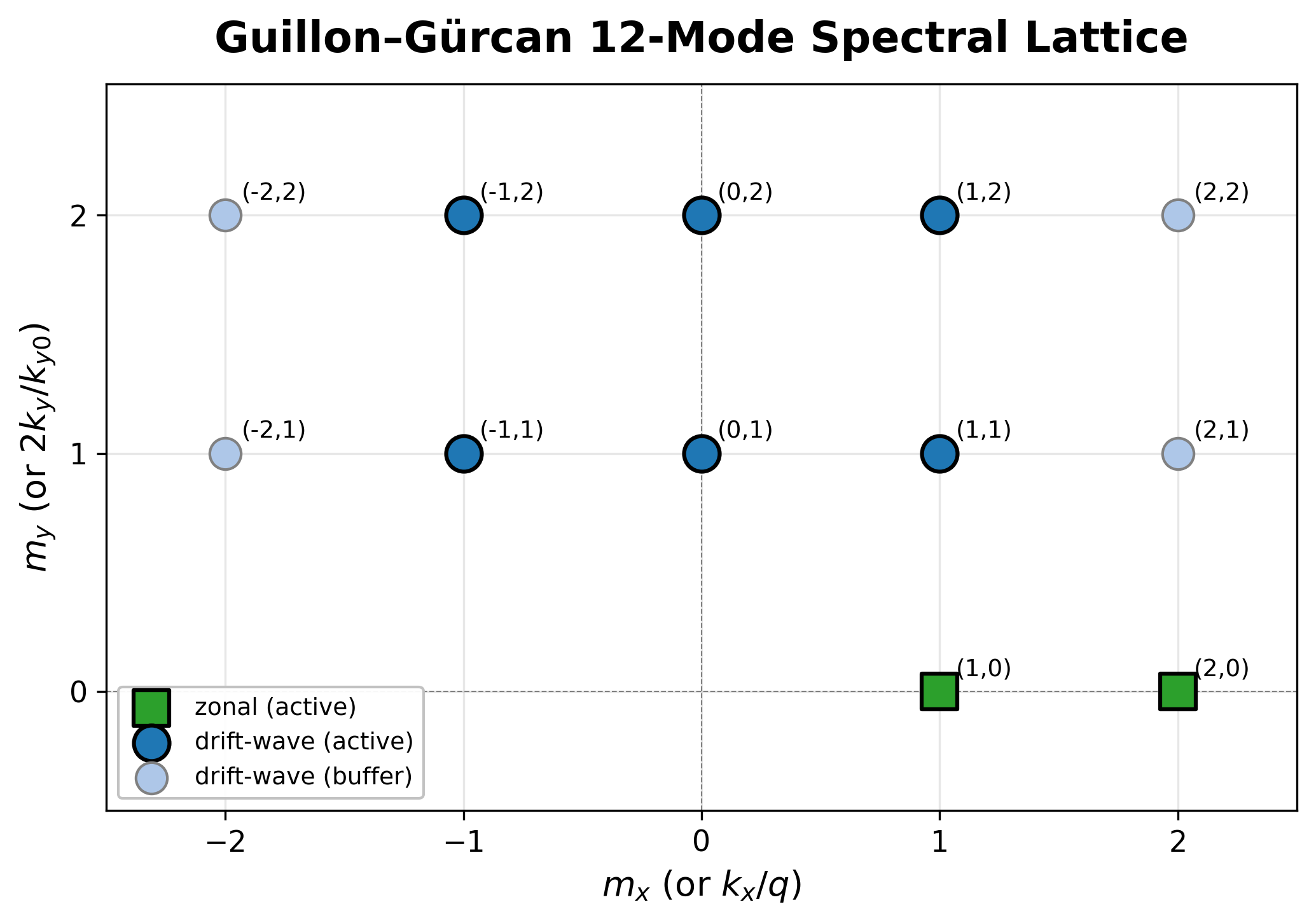}
\caption{Canonical 12-mode Fourier truncation, showing the twelve independent retained wavevectors. Conjugate partners are supplied by the reality condition. Zonal modes have $k_y=0$, while drift-wave modes have $k_y\neq0$. The four outer sidebands remain dynamically active but receive additional damping as buffer modes.}
\label{fig:lattice}
\end{figure*}

The four outer sidebands $(\pm2,1)$ and $(\pm2,2)$ act as buffer modes. They remain resolved and participate in the retained nonlinear triads, but receive the additional damping specified in Section~\ref{sec:evaluation_protocol}. Figure~\ref{fig:lattice} shows the retained lattice.

The full MHW system exhibits a transition between turbulent and ZF-dominated states~\cite{numata2007}. The 12-mode truncation retains this first-order-like transition, characterized by $C/\kappa$ and the zonal fraction as an effective order parameter~\cite{guillon2025}. We hold $\kappa$ fixed and sweep $C$ upward and downward to determine the transition thresholds and the hysteresis width.

We diagnose the transition using the zonal kinetic-energy fraction
\begin{equation}
\Xi_K=\frac{K_{\mathrm{ZF}}}{K_{\mathrm{ZF}}+K_{\mathrm{DW}}},
\label{eq:XiK}
\end{equation}
where
\[
K_{\mathrm{ZF}}=\frac12\sum_{k_y=0}k^2|\phi_{\mathbf k}|^2,
\qquad K_{\mathrm{DW}}=\frac12\sum_{k_y\neq0}k^2|\phi_{\mathbf k}|^2.
\]
The sums extend over the full Fourier spectrum for DNS and over $\mathcal R$ for the ROM. Thus $\Xi_K$ measures the zonal share of kinetic energy within each model's spectrum. Agreement in this diagnostic does not by itself imply agreement in the energy distribution among the retained modes. Let $C_\uparrow$ and $C_\downarrow$ denote the transition thresholds on the increasing and decreasing sweeps. The hysteresis width is
\begin{equation}
\Delta C=C_\uparrow-C_\downarrow.
\label{eq:hysteresis_width_background}
\end{equation}
The thresholds are estimated from the first crossing of $\Xi_K=0.5$ on each sweep, using the averaging and interpolation procedures in Section~\ref{sec:evaluation_protocol}. This definition characterizes hysteresis under the prescribed finite-duration sweep. The unclosed truncation retains nonlinear interactions within $\mathcal R$ but omits feedback involving $\mathcal U$. Section~\ref{sec:methodology} develops a closure for this missing feedback, and Section~\ref{sec:results} evaluates its effect on the transition thresholds and hysteresis width.


\section{Methodology}
\label{sec:methodology}

The closure represents the nonlinear feedback of discarded modes on the retained system. Inspired by the Mori--Zwanzig framework, we first characterize their coupling to the retained dynamics. We then use the resulting structure to construct a low-dimensional phenomenological closure.                                     

\subsection{Resolved--unresolved partition}
\label{sec:mz_split}

We partition the spectral state as $\mathbf u=(\mathbf u^{\mathcal R},\mathbf u^{\mathcal U})$ using the mode sets defined in Section~\ref{sec:background}. The MZ formalism expresses the influence of the unresolved state \(\mathbf u^{\mathcal U}\) on the resolved dynamics through projected dynamics, memory and orthogonal dynamics~\cite{nakajima1958,zwanzig1961,mori1965,chorin2000,givon2004}. For $\dot{\mathbf u}=\mathbf F(\mathbf u)$, the Liouville operator
$\mathcal L=\mathbf F\cdot\nabla_{\mathbf u}$ generates the evolution
of observables along the full dynamics. Let $\mathcal P$ project onto
observables of the retained state, and let
$\mathcal Q=1-\mathcal P$ denote the complementary projection. Thus
$e^{t\mathcal L}\mathbf u_{\mathbf k}$ is the mode observable evaluated
after evolution for time $t$. The exact resolved evolution is

\begin{equation}
\begin{aligned}
\frac{d}{dt}e^{t\mathcal L}\mathbf u_{\mathbf k}
&=e^{t\mathcal L}\mathcal P\mathcal L\mathbf u_{\mathbf k}
+e^{t\mathcal Q\mathcal L}\mathcal Q\mathcal L\mathbf u_{\mathbf k}\\
&\quad+\int_0^t e^{(t-\tau)\mathcal L}\mathcal P\mathcal L
 e^{\tau\mathcal Q\mathcal L}\mathcal Q\mathcal L\mathbf u_{\mathbf k}\,d\tau,\\
&\hspace{5em}\mathbf k\in\mathcal R.
\end{aligned}
\label{eq:mz_gle}
\end{equation}
The terms on the right-hand side describe projected dynamics, orthogonal dynamics, and memory, respectively. The projected dynamics need not coincide with the unclosed truncation because they can contain averaged unresolved contributions. We use this decomposition only to guide the closure.

The linear operator acts independently at each wavevector, so coupling between $\mathcal R$ and $\mathcal U$ occurs only through the advective triads $\mathbf p+\mathbf q=\mathbf k$. Because the $E\times B$ drift velocity is divergence free, the nonlinear advection term redistributes energy across scales. It cannot change the total energy. Therefore,

\begin{equation}
\left.
\frac{d}{dt}
\sum_{\mathbf k\ne\mathbf0}
\left(
\frac{|\Omega_{\mathbf k}|^2}{k^2}
+
|n_{\mathbf k}|^2
\right)
\right|_{\mathrm{nl}}
=0.
\label{eq:nonlinear_energy_conservation}
\end{equation}

For later use, we write this energy as
\begin{equation}
E
=
\langle\mathbf u,\mathbf u\rangle_{\mathbf G},
\qquad
\langle\mathbf v,\mathbf w\rangle_{\mathbf G}
\equiv
\sum_{\mathbf k\ne\mathbf0}
\mathbf v_{\mathbf k}^{\dagger}
\mathbf G_{\mathbf k}
\mathbf w_{\mathbf k},
\qquad
\mathbf G_{\mathbf k}
=
\operatorname{diag}(1/k^2,1),
\label{eq:energy_metric}
\end{equation}
where the conventional factor of $1/2$ is omitted.

In the transformed coordinates,
\[
\widetilde{\mathbf G}_{\mathbf k}
=
\mathbf V_{\mathbf k}^{\dagger}
\mathbf G_{\mathbf k}
\mathbf V_{\mathbf k}
=
\operatorname{diag}(w_+,w_-),
\qquad
w_+=\frac{1+k^2}{k^2},
\qquad
w_-=\frac{1+k^2}{k^4}.
\]
The modal energy is therefore
\[
E_{\mathbf k}
=
w_+|\widetilde u_{+,\mathbf k}|^2
+
w_-|\widetilde u_{-,\mathbf k}|^2.
\]
We define the transformed inner product and norm by
\[
\left\langle
\widetilde{\mathbf v},
\widetilde{\mathbf w}
\right\rangle_{\widetilde{\mathbf G}}
\equiv
\sum_{\mathbf k}
\widetilde{\mathbf v}_{\mathbf k}^{\dagger}
\widetilde{\mathbf G}_{\mathbf k}
\widetilde{\mathbf w}_{\mathbf k},
\qquad
\|\widetilde{\mathbf v}\|_{\widetilde{\mathbf G}}^2
\equiv
\left\langle
\widetilde{\mathbf v},
\widetilde{\mathbf v}
\right\rangle_{\widetilde{\mathbf G}},
\]
where the sum is restricted to the indicated mode set.

The same inner product gives the energy rate,
\begin{equation}
\dot E
=
2\operatorname{Re}
\langle
\widetilde{\mathbf u},
\dot{\widetilde{\mathbf u}}
\rangle_{\widetilde{\mathbf G}}.
\label{eq:energy_power_pairing}
\end{equation}

\subsection{Energy transfer between resolved and unresolved modes}
\label{sec:ledger}

The energy separates into resolved and unresolved contributions,
$E=E_{\mathcal R}+E_{\mathcal U}$, where
\[
E_{\mathcal S}\equiv
\sum_{\mathbf k\in\mathcal S}
\left(
\frac{|\Omega_{\mathbf k}|^2}{k^2}
+
|n_{\mathbf k}|^2
\right),
\qquad
\mathcal S\in\{\mathcal R,\mathcal U\}.
\]
Here $E_{\mathcal U}$ is the exact energy of the unresolved spectrum. The reservoir introduced below provides a reduced representation of its activity.

We define the nonlinear transfer rate by

\begin{equation}
\Pi_{\mathcal R\rightarrow\mathcal U}
\equiv
-\left.\dot E_{\mathcal R}\right|_{\mathrm{nl}}
=
\left.\dot E_{\mathcal U}\right|_{\mathrm{nl}},
\label{eq:flux_def}
\end{equation}
where $\Pi_{\mathcal R\rightarrow\mathcal U}>0$ denotes transfer into the unresolved spectrum. Conservation does not constrain the sign of this transfer.

Including the linear contributions gives the exact balances
\begin{equation}
\begin{aligned}
\dot E_{\mathcal R}
&=
\dot E_{\mathcal R}^{\mathrm{lin}}
-\Pi_{\mathcal R\rightarrow\mathcal U},\\
\dot E_{\mathcal U}
&=
\dot E_{\mathcal U}^{\mathrm{lin}}
+\Pi_{\mathcal R\rightarrow\mathcal U}.
\end{aligned}
\label{eq:ledger}
\end{equation}
For either mode set,
\[
\dot E_{\mathcal S}^{\mathrm{lin}}
=
-2\operatorname{Re}
\sum_{\mathbf k\in\mathcal S}
\mathbf u_{\mathbf k}^{\dagger}
\mathbf G_{\mathbf k}\mathbf L_{\mathbf k}\mathbf u_{\mathbf k}.
\]
This contribution includes both background-gradient drive and dissipation and need not be negative. These balances guide the reservoir and feedback models below, but do not by themselves determine their functional forms.

\subsection{Closure construction}
\label{sec:closure}
The truncation retains interactions among the resolved modes but omits
their coupling to the unresolved spectrum. We coarse-grain the unresolved energy into a reservoir \(\mathcal E(t)\), then model its evolution and feedback on the retained modes.

\subsubsection{Reservoir dynamics}

Equation~\eqref{eq:ledger} separates linear evolution from nonlinear
energy transfer between the resolved and unresolved modes. To identify
the interactions responsible for this transfer, we write the MHW
dynamics in the two mode sets:
\begin{equation}
\begin{aligned}
\dot{\widetilde{\mathbf u}}^{\mathcal R}
+\mathbf M_{\mathcal R}\widetilde{\mathbf u}^{\mathcal R}
&=\mathbf c^{\mathrm{RRR}}
+\mathbf c^{\mathrm{RUR}}
+\mathbf c^{\mathrm{UUR}},\\
\dot{\widetilde{\mathbf u}}^{\mathcal U}
+\mathbf M_{\mathcal U}\widetilde{\mathbf u}^{\mathcal U}
&=\mathbf c^{\mathrm{RRU}}
+\mathbf c^{\mathrm{RUU}}
+\mathbf c^{\mathrm{UUU}}.
\end{aligned}
\label{eq:split_dynamics}
\end{equation}
The three letters identify the two input mode sets and the output mode
set; mixed channels include both input orderings.

The resolved zonal modes evolve through nonlinear coupling, while the
resolved DW modes also experience linear drive and dissipation.
Separating $k_y=0$ and $k_y\neq0$ gives
\begin{equation}
\begin{aligned}
\dot{\widetilde{\mathbf u}}_Z
&=\mathbf c_Z^{\mathrm{RRR}}
+\mathbf c_Z^{\mathrm{RUR}}
+\mathbf c_Z^{\mathrm{UUR}},\\
\dot{\widetilde{\mathbf u}}_{\mathrm{DW}}
+\mathbf M_{\mathrm{DW}}\widetilde{\mathbf u}_{\mathrm{DW}}
&=\mathbf c_{\mathrm{DW}}^{\mathrm{RRR}}
+\mathbf c_{\mathrm{DW}}^{\mathrm{RUR}}
+\mathbf c_{\mathrm{DW}}^{\mathrm{UUR}}.
\end{aligned}
\label{eq:resolved_zonal_dw_split}
\end{equation}

Linear drive and dissipation vary across the unresolved modes, so their
net contribution depends on how $E_{\mathcal U}$ is distributed among
them. Restricting the energy-rate identity in
Eq.~\eqref{eq:energy_power_pairing} to the unresolved modes and their
linear dynamics, we express the net linear power as an instantaneous
rate multiplying the unresolved energy:
\[
\dot E_{\mathcal U}^{\mathrm{lin}}
=
-\gamma_{\mathrm{lin}}(t)E_{\mathcal U},
\qquad
\gamma_{\mathrm{lin}}(t)
\equiv
\frac{
2\operatorname{Re}
\left\langle
\widetilde{\mathbf u}^{\mathcal U},
\mathbf M_{\mathcal U}\widetilde{\mathbf u}^{\mathcal U}
\right\rangle_{\widetilde{\mathbf G}}
}{
E_{\mathcal U}
}.
\]
Positive $\gamma_{\mathrm{lin}}$ denotes net linear loss and negative
$\gamma_{\mathrm{lin}}$ net linear drive. Under the reservoir
coarse-graining $E_{\mathcal U}\rightarrow\mathcal E$, we approximate
this composition-dependent rate by a constant $\gamma_c^{(0)}$:
\begin{equation}
\left.\dot{\mathcal E}\right|_{\mathrm{lin}}
=
-\gamma_c^{(0)}\mathcal E.
\label{eq:reservoir_linear_model}
\end{equation}

We now consider the nonlinear contribution to the reservoir dynamics.
Interactions entirely within $\mathcal U$ redistribute energy among
unresolved modes but do not change the total unresolved energy.
Exchange between $\mathcal R$ and $\mathcal U$ therefore enters through
the RRU and RUU channels.

The RRU channel contains two resolved inputs and forces the unresolved
modes. Under a short-memory, or Markovian, approximation, we model the
unresolved response as local in time, so that its contribution to the
reservoir energy is proportional to the squared instantaneous RRU
forcing~\cite{orszag1970,bowman1993}.

Since ZF--ZF interactions vanish by Eq.~\eqref{eq:Nk}, the RRU forcing
contains DW--DW and mixed ZF--DW contributions,
\[
\mathbf c^{\mathrm{RRU}}
=
\mathbf c_{\mathrm{DW-DW}}^{\mathrm{RRU}}
+
\mathbf c_{\mathrm{ZF-DW}}^{\mathrm{RRU}}.
\]
As a minimal source model, we retain the DW--DW contribution. Because
this forcing is quadratic in the resolved DW amplitudes,
\[
\left\|
\mathbf c_{\mathrm{DW-DW}}^{\mathrm{RRU}}
\right\|_{\widetilde{\mathbf G}}^2
\sim
\bigl(E_{\mathrm{DW}}^{\mathcal R}\bigr)^2,
\qquad
E_{\mathrm{DW}}^{\mathcal R}
=
\sum_{\substack{\mathbf k\in\mathcal R\\k_y\neq0}}
E_{\mathbf k}.
\]
Absorbing the response coefficient and reservoir normalization into
$\gamma_{\mathrm{in}}>0$ gives
\begin{equation}
S_{\mathrm{RRU}}
=
\gamma_{\mathrm{in}}
\bigl(E_{\mathrm{DW}}^{\mathcal R}\bigr)^2.
\label{eq:RRU_source_model}
\end{equation}
This source represents reservoir excitation by resolved DW--DW
interactions; modulation by mixed ZF--DW interactions is not included.

The RUU channel contains one resolved and one unresolved input. In the
unresolved-energy rate, $\mathbf c^{\mathrm{RUU}}$ is paired with the
unresolved state, so the contribution contains two unresolved
amplitudes and is therefore proportional to $E_{\mathcal U}$ at fixed
resolved state. We represent the remaining
state dependence by a coarse-grained rate $\gamma_{\mathrm{RUU}}$.
Under $E_{\mathcal U}\rightarrow\mathcal E$,
\[
S_{\mathrm{RUU}}
=
\gamma_{\mathrm{RUU}}\mathcal E,
\]
where $\gamma_{\mathrm{RUU}}$ may have either sign.

Combining this exchange with the unresolved linear contribution
$-\gamma_c^{(0)}\mathcal E$ gives the net relaxation rate
\[
\gamma_c
\equiv
\gamma_c^{(0)}-\gamma_{\mathrm{RUU}}>0.
\]
The reservoir dynamics are therefore
\begin{equation}
\dot{\mathcal E}
=
\gamma_{\mathrm{in}}
\bigl(E_{\mathrm{DW}}^{\mathcal R}\bigr)^2
-
\gamma_c\mathcal E.
\label{eq:reservoir_model}
\end{equation}
Thus, resolved DW--DW interactions supply the reservoir, while the
combined unresolved linear dynamics and RUU exchange determine the
relaxation time $\gamma_c^{-1}$.

The reservoir therefore provides a coarse-grained representation of the
energy in the unresolved spectrum. Having modeled the evolution of the reservoir, we next determine the coupling between the reservoir and the resolved modes, beginning with the zonal flow.

\subsubsection{Zonal-flow feedback}

The resolved zonal modes couple to the unresolved spectrum through RUR
and UUR interactions in Eq.~\eqref{eq:resolved_zonal_dw_split}, with
\[
\|\mathbf c_Z^{\mathrm{RUR}}\|_{\widetilde{\mathbf G}}
\sim
\|\widetilde{\mathbf u}^{\mathcal R}\|_{\widetilde{\mathbf G}}
\|\widetilde{\mathbf u}^{\mathcal U}\|_{\widetilde{\mathbf G}},
\qquad
\|\mathbf c_Z^{\mathrm{UUR}}\|_{\widetilde{\mathbf G}}
\sim
\|\widetilde{\mathbf u}^{\mathcal U}\|_{\widetilde{\mathbf G}}^2.
\]
In the saturated regime, we assume that the unresolved energy dominates
the resolved energy,
\begin{equation}
\varepsilon
\equiv
\frac{\|\widetilde{\mathbf u}^{\mathcal U}\|_{\widetilde{\mathbf G}}}
{\|\widetilde{\mathbf u}^{\mathcal R}\|_{\widetilde{\mathbf G}}}
=
\sqrt{\frac{E_{\mathcal U}}{E_{\mathcal R}}}
\gg1.
\label{eq:epsilon_def}
\end{equation}
In this regime, interactions involving two unresolved
modes are expected to dominate those involving only one. We therefore
retain UUR as the leading unresolved contribution to the zonal dynamics. From Eq.~\eqref{eq:resolved_zonal_dw_split} and Eq.~\eqref{eq:epsilon_def}, we have
\[
\dot{\widetilde{\mathbf u}}_Z
\simeq
\mathbf c_Z^{\mathrm{RRR}}
+
\mathbf c_Z^{\mathrm{UUR}}.
\]

The instantaneous UUR feedback depends on the detailed configuration of the unresolved
modes discarded by the truncation. Following the statistical-projection
interpretation of Mori--Zwanzig reduction~\cite{chorin_physd2002}, we characterize their
collective effect by averaging over unresolved configurations at fixed
resolved state and unresolved energy,
\begin{equation}
\mathbf f_Z^{\mathrm{UUR}}
\equiv
\left\langle
\mathbf c_Z^{\mathrm{UUR}}
\right\rangle_{
\widetilde{\mathbf u}^{\mathcal R},\,E_{\mathcal U}} ,
\label{eq:zonal_ensemble_average}
\end{equation}
where the subscript denotes the quantities, resolved state and unresolved energy respectively, held fixed in the ensemble
average. The zonal dynamics are then approximated by
\[
\dot{\widetilde{\mathbf u}}_Z
\simeq
\mathbf c_Z^{\mathrm{RRR}}
+
\mathbf f_Z^{\mathrm{UUR}}.
\]

The contribution of $\mathbf f_Z^{\mathrm{UUR}}$ to the zonal-energy
rate is
$2\operatorname{Re}
\langle\widetilde{\mathbf u}_Z,
\mathbf f_Z^{\mathrm{UUR}}\rangle_{\widetilde{\mathbf G}}$.
We isolate this energy-changing component by projecting onto the
zonal state using the real part of the energy inner product:
\begin{equation}
\mathbf f_{Z,\parallel}^{\mathrm{UUR}}
=
\frac{
\operatorname{Re}
\langle\widetilde{\mathbf u}_Z,
\mathbf f_Z^{\mathrm{UUR}}\rangle_{\widetilde{\mathbf G}}
}{
E_Z^{\mathcal R}
}
\widetilde{\mathbf u}_Z
\equiv
-\nu_{\mathrm{eff}}\widetilde{\mathbf u}_Z,
\qquad E_Z^{\mathcal R}>0,
\label{eq:zonal_parallel_feedback}
\end{equation}
where
$E_Z^{\mathcal R}
=\sum_{\substack{\mathbf k\in\mathcal R\\k_y=0}}E_{\mathbf k}
=\|\widetilde{\mathbf u}_Z\|_{\widetilde{\mathbf G}}^2$
is the total retained zonal energy.
The remaining component makes no instantaneous contribution to the
zonal-energy change and is neglected in the scalar closure.

The projection does not constrain the sign of $\nu_{\mathrm{eff}}$.
Separating the unresolved-energy factor from the remaining dependence
on the resolved state and unresolved statistics, we write
\begin{equation}
\nu_{\mathrm{eff}}=E_{\mathcal U}\alpha_Z.
\end{equation} 

Under the coarse-graining \(E_{\mathcal U}\rightarrow\mathcal E\), we neglect the remaining state dependence and represent \(\alpha_Z\) by a constant positive coefficient, consistent with the use of effective relaxation rates in Markovian turbulence closures~\cite{orszag1970,bowman1993}. Absorbing the reservoir normalization into \(\nu_{\mathrm{ZF}}\) then gives
$\nu_{\mathrm{eff}}\simeq\nu_{\mathrm{ZF}}\mathcal E$.

The zonal closure is therefore
\begin{equation}
\dot{\widetilde{\mathbf u}}_Z
=
\mathbf c_Z^{\mathrm{RRR}}
-
\nu_{\mathrm{ZF}}\mathcal E\,
\widetilde{\mathbf u}_Z,
\qquad
\nu_{\mathrm{ZF}}>0.
\label{eq:zonal_scalar_closure}
\end{equation}

The zonal closure therefore represents the coarse-grained unresolved feedback through a reservoir-dependent damping of the retained zonal modes. We next construct the corresponding feedback on the DW modes, focusing on the ZF-dominated plateau of the hysteresis loop, where DW perturbations are weak.

\subsubsection{Drift-wave feedback}
\label{sec:dw_feedback}

On this plateau, the zonal field remains strong while the DW
amplitudes are small,
\begin{equation}
\|\widetilde{\mathbf u}_{\mathrm{DW}}\|_{\widetilde{\mathbf G}}
\ll
\|\widetilde{\mathbf u}_{Z}\|_{\widetilde{\mathbf G}}.
\label{eq:DW_small_perturbation}
\end{equation}
The zonal field can therefore couple a resolved DW perturbation into the
unresolved spectrum, whose response subsequently couples back to the
resolved DW modes. We construct this RRU--RUR return by retaining terms
linear in the DW perturbation. Since ZF--ZF interactions vanish and
DW--DW forcing is quadratic in the perturbation, the leading RRU forcing
contains one zonal and one DW amplitude:
\[
\begin{aligned}
\mathbf c^{\mathrm{RRU}}
&=
\mathbf c^{\mathrm{RRU}}
[\widetilde{\mathbf u}_{Z},\widetilde{\mathbf u}_{\mathrm{DW}}]
+
\mathbf c^{\mathrm{RRU}}
[\widetilde{\mathbf u}_{\mathrm{DW}},\widetilde{\mathbf u}_{Z}]
\\
&\quad+
O\!\left(
\|\widetilde{\mathbf u}_{\mathrm{DW}}\|_{\widetilde{\mathbf G}}^2
\right).
\end{aligned}
\]

Thus, at fixed zonal field, $\widetilde{\mathbf u}_Z$, the leading RRU
forcing is linear in the DW perturbation, $\widetilde{\mathbf u}_{\mathrm{DW}}$. The resulting unresolved
perturbations evolve under their linear dynamics and further
interactions with the zonal field, before returning to the
resolved DW modes through the RUR channel. We denote this
feedback by $\mathbf f_{\mathrm{DW}}^{\mathrm{RUR}}$.

For example,
\[
(2,1)
\xrightarrow{\,(1,0)\,}
(3,1)\in\mathcal U
\xrightarrow{\,(-1,0)\,}
(2,1)\in\mathcal R .
\]

For a first-harmonic zonal field of fixed
structure and complex amplitude $Z_1$, the two interactions
in this direct return contribute one factor of $Z_1$ and
one of $Z_1^*$. Under the short-memory approximation, both
are evaluated at the current time, giving
$Z_1Z_1^*=|Z_1|^2$. We define the first-harmonic zonal
energy by
\begin{equation}
E_{Z,1}
\equiv
\sum_{\substack{\mathbf k\in\mathcal R\\
k_y=0,\ |k_x|=q}}
E_{\mathbf k}
\propto |Z_1|^2,
\label{eq:E1_def}
\end{equation}
including both conjugate partners.
This leading return motivates a feedback proportional to
$E_{Z,1}\widetilde{\mathbf u}_{\mathrm{DW}}$.
As a modeling choice, we apply this dependence to all
retained DW modes and represent the feedback by
instantaneous scalar damping:
\begin{equation}
\mathbf f_{\mathrm{DW}}^{\mathrm{RUR}}
\simeq
-\eta_{\mathrm{DW}}
E_{Z,1}
\widetilde{\mathbf u}_{\mathrm{DW}},
\qquad
\eta_{\mathrm{DW}}\geq0 .
\label{eq:DW_scalar_return}
\end{equation}
Together with the reservoir dynamics and zonal feedback above, this
completes the three closure terms.

\subsubsection{The closure}
\label{sec:closed_system}

The resulting closure combines reservoir excitation and relaxation with
feedback on the retained zonal and DW modes. The closure takes the form
\begin{equation}
\begin{aligned}
\dot{\widetilde{\mathbf u}}_Z
&=
\mathbf c_Z^{\mathrm{RRR}}
-\nu_{\mathrm{ZF}}\mathcal E\,\widetilde{\mathbf u}_Z,
\\
\dot{\widetilde{\mathbf u}}_{\mathrm{DW}}
&=
-\mathbf M_{\mathrm{DW}}\widetilde{\mathbf u}_{\mathrm{DW}}
+\mathbf c_{\mathrm{DW}}^{\mathrm{RRR}}
-\eta_{\mathrm{DW}}
E_{Z,1}
\widetilde{\mathbf u}_{\mathrm{DW}},
\\
\dot{\mathcal E}
&=
\gamma_{\mathrm{in}}
\bigl(E_{\mathrm{DW}}^{\mathcal R}\bigr)^2
-\gamma_c\mathcal E .
\end{aligned}
\label{eq:closure_physical}
\end{equation}

The reservoir equation can be solved explicitly as
\[
\mathcal E(t)
=
e^{-\gamma_c t}\mathcal E(0)
+
\gamma_{\mathrm{in}}
\int_0^t
e^{-\gamma_c(t-s)}
\bigl(E_{\mathrm{DW}}^{\mathcal R}(s)\bigr)^2\,ds .
\]
The first term carries the contribution of the initial reservoir state. The integral contains an exponentially weighted history of the resolved DW activity: activity at the earlier time $s$ is weighted by $e^{-\gamma_c(t-s)}$. Substituting this expression into the zonal-feedback term in Eq.~\eqref{eq:closure_physical} gives
\[
-\nu_{\mathrm{ZF}}\mathcal E(t)\widetilde{\mathbf u}_Z(t)
=
-\nu_{\mathrm{ZF}}e^{-\gamma_c t}\mathcal E(0)
\widetilde{\mathbf u}_Z(t)
-
\widetilde{\mathbf u}_Z(t)
\int_0^t
K(t-s)
\bigl(E_{\mathrm{DW}}^{\mathcal R}(s)\bigr)^2\,ds ,
\]
where the modeled memory kernel is
\[
K(\tau)
=
\nu_{\mathrm{ZF}}\gamma_{\mathrm{in}}e^{-\gamma_c\tau},
\qquad \tau\geq0 .
\]
Thus, past DW activity modifies the present zonal damping, with older contributions progressively forgotten on the characteristic timescale $\gamma_c^{-1}$.

For calibration and integration, it is convenient to combine the reservoir normalization and zonal-feedback strength into parameters that enter the resolved dynamics directly. We absorb the zonal-feedback coefficient into the reservoir coordinate,
\begin{equation}
W\equiv\nu_{\mathrm{ZF}}\mathcal E,
\qquad
\Lambda\equiv
\frac{\nu_{\mathrm{ZF}}\gamma_{\mathrm{in}}}{\gamma_c}.
\label{eq:W_Lambda_def}
\end{equation}

The closed reduced-order model is
\begin{equation}
\begin{aligned}
E_{\mathrm{DW}}^{\mathcal R}
&=
\sum_{\substack{\mathbf k\in\mathcal R\\k_y\neq0}}E_{\mathbf k},
\\
E_{Z,1}
&=
\sum_{\substack{\mathbf k\in\mathcal R\\k_y=0,\ |k_x|=q}}E_{\mathbf k},
\\
\dot W
&=
\gamma_c\left[
\Lambda\bigl(E_{\mathrm{DW}}^{\mathcal R}\bigr)^2-W
\right],
\\
W_{\mathbf k}
&=
\begin{cases}
W, & k_y=0,\\[1mm]
\displaystyle
\eta_{\mathrm{DW}}
E_{Z,1},
& k_y\neq0,
\end{cases}
\\
\dot{\widetilde{\mathbf u}}_{\mathbf k}
\hwromtwocolumn{\\&}{&}=
-
\begin{pmatrix}
\hwromtwocolumn{%
\lambda_+-i\omega_{\mathbf k}+W_{\mathbf k}
&
-i\omega_{\mathbf k}
\\
i\omega_{\mathbf k}
&
}{%
\lambda_+-i\omega_{\mathbf k}+W_{\mathbf k} & -i\omega_{\mathbf k}\\
i\omega_{\mathbf k} &
}
\lambda_-+i\omega_{\mathbf k}+W_{\mathbf k}
\end{pmatrix}
\widetilde{\mathbf u}_{\mathbf k}
\hwromtwocolumn{\\&\quad}{}
+
\mathbf c_{\mathbf k}^{\mathrm{RRR}},
\qquad
\mathbf k\in\mathcal R .
\end{aligned}
\label{eq:complete_closure_W}
\end{equation}

The change of variable $W=\nu_{\mathrm{ZF}}\mathcal E$ writes the reservoir contribution to the zonal equation as $-W\widetilde{\mathbf u}_Z$. In this form, the modeled kernel is
\[
K(\tau)=\Lambda\gamma_c e^{-\gamma_c\tau},
\]
and its integrated strength is
\[
\int_0^\infty K(\tau)\,d\tau=\Lambda .
\]
Hence, $\Lambda$ sets the overall strength of the reservoir-mediated memory, while $\gamma_c^{-1}$ sets its decay time.

\subsection{Closure parameters}
\label{sec:closure_parameters}

We fix the reservoir relaxation rate at $\gamma_c=1$ and calibrate $\Lambda$ to the DNS hysteresis-loop area. The closure parameters are
\begin{equation}
\gamma_c=1,
\qquad
\Lambda=7.5\times10^{-5},
\qquad
\eta_{\mathrm{DW}}=3.75\times10^{-3}.
\label{eq:closure_parameters}
\end{equation}
Appendix~\ref{app:benchmark} gives their selection procedure.
Here $\eta_{\mathrm{DW}}$ sets the DW damping per unit first-harmonic zonal energy.

\subsection{Evaluation}
\label{sec:evaluation_protocol}
The ROM lattice and buffer-damping prescription follow Guillon and G\"urcan~\cite{guillon2025}. The lattice remains fixed at its $C=0.1$ geometry, with $q=0.6\,k_{y0}(0.1)$, so varying $C$ changes the linear rates without changing the retained wavevectors. Ordinary viscosity is set to zero in the ROM, and the four outer sidebands receive the additional damping
\begin{equation}
d_{\mathrm{buf}}(C)
=
10\sqrt{\gamma_{\max}(C)\,\omega_{r,\max}(C)},
\label{eq:buffer_damping}
\end{equation}
where $\gamma_{\max}(C)$ is the maximum growth rate of the inviscid homogeneous MHW system at $\kappa=1$ and $k_x=0$. The quantity $\omega_{r,\max}(C)$ is the real frequency of the corresponding mode on the positive-$k_y$ branch. DNS uses the viscosity prescription in Eq.~\eqref{eq:nu_dns}.

The control parameter is swept from $C=0.05$ to $0.20$ and then down to $0.04$ in steps of $0.01$, giving $32$ plateaus with the turning point counted once. Each plateau lasts $200/\gamma_{\max}(C)$, and its final state initializes the next plateau. The ROM uses fixed-step fourth-order Runge--Kutta integration with $\Delta t=0.1$ and an initial warmup of $400/\gamma_{\max}(0.05)$. DNS uses adaptive integration and a warmup of $200/\gamma_{\max}(0.05)$. All ROM runs use $\kappa=1$, and the reservoir is initialized with $\mathcal E(0)=0$ before the warmup.

For the hysteresis comparison, the ROM uses second-half plateau averages of $\Xi_K$ and DNS uses full-plateau averages. The particle-flux comparison uses the common averaging window $100<\gamma_{\max}(C)(t-t_{\mathrm{step}})<200$ for DNS and both reduced models, where $t_{\mathrm{step}}$ is the start of the plateau.

For each realization, $C_\uparrow$ and $C_\downarrow$ are determined from the first strict crossing of $\Xi_K=0.5$ along the corresponding sweep, with linear interpolation between adjacent plateau averages. The width $\Delta C=C_\uparrow-C_\downarrow$ is calculated for each realization with both crossings before ensemble averaging. Transition thresholds are averaged over realizations with the corresponding crossing. Because the reduced models use matching initial conditions, comparisons between them are paired realization by realization. Appendix~\ref{app:benchmark} gives the numerical details, the parameter selection and the construction of the DNS diagnostics.
\section{Results}
\label{sec:results}

We compare the unclosed 12-mode truncation and the closed ROM with full-field DNS. The two reduced models share the same 32 initial conditions, allowing paired comparisons. The $512\times512$ DNS reference uses an independent ensemble of 32 realizations. The evaluation is defined in Section~\ref{sec:evaluation_protocol}.

\subsection{Recovery of the hysteretic transition}
\label{sec:hyst_match}

\begin{figure*}[htbp]
\centering
\includegraphics[width=\textwidth]{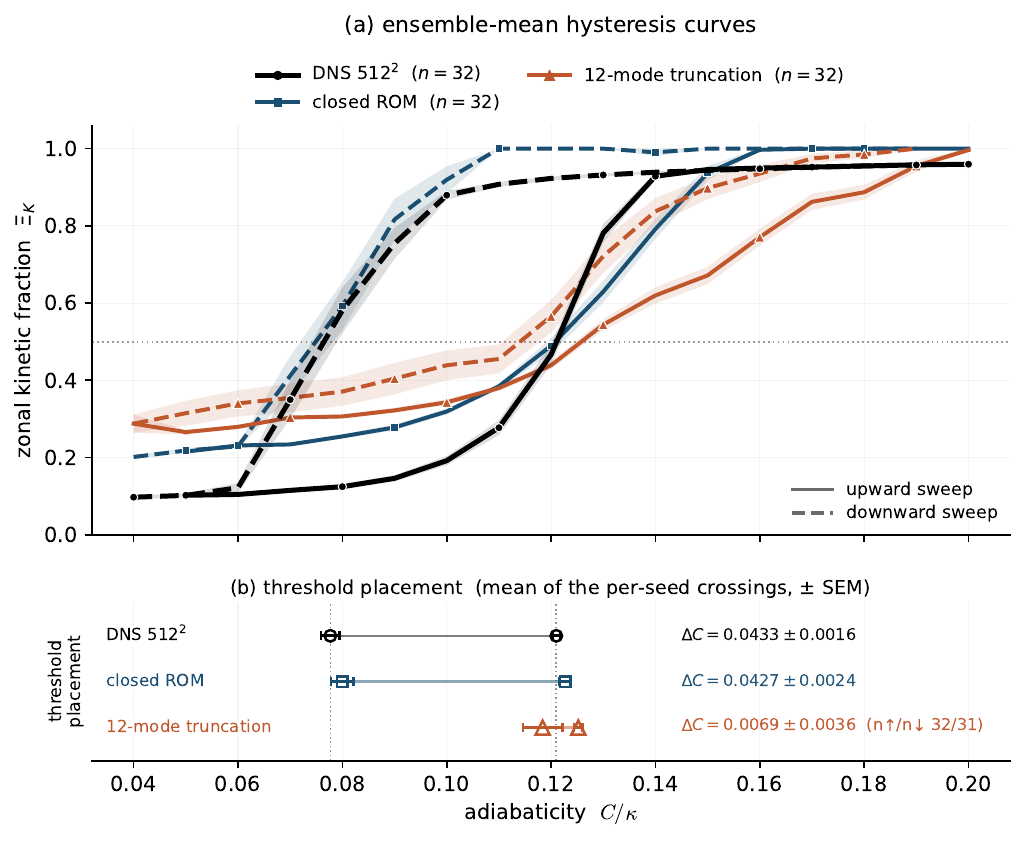}
\caption{(a) Ensemble-averaged zonal kinetic-energy fraction $\Xi_K$ on the increasing- and decreasing-$C$ sweeps for $512\times512$ DNS, the 12-mode truncation, and the closed ROM. Each ensemble contains 32 realizations; the two reduced models use the same initial conditions. Shading is defined in Appendix~B. (b) Transition thresholds obtained from the $\Xi_K=0.5$ crossings. Error bars denote one standard error. The hysteresis widths are $0.0433\pm0.0016$ for DNS, $0.0069\pm0.0036$ for the truncation, and $0.0427\pm0.0024$ for the closed ROM. The truncation width is based on the 31 realizations with both crossings.}
\label{fig:headline}
\end{figure*}

The closure recovers most of the hysteresis lost under spectral truncation, as shown in Fig.~\ref{fig:headline} and Table~\ref{tab:transition_results}. The closure increases the hysteresis width from
$\Delta C=0.0069\pm0.0036$ in the unclosed model to $0.0427\pm0.0024$, compared
with $0.0433\pm0.0016$ in DNS, a gain of $0.0358$, or about $98\%$ of the
truncation--DNS difference. The corresponding threshold shifts are
reported in Table~\ref{tab:transition_results}. Fourteen of the $32$ unclosed realizations produce inverted loops, with $C_\uparrow<C_\downarrow$. No inverted loops occur in the closed ROM or DNS ensembles.

The closed-ROM thresholds lie slightly above DNS, by \(0.0017\) for \(C_\uparrow\) and \(0.0023\) for \(C_\downarrow\). These offsets are of the same order as the reported standard errors. The hysteresis width differs from DNS by only \(0.0006\).

\begin{table*}[htbp]
\centering
\caption{Mean transition thresholds and hysteresis widths obtained from the first $\Xi_K=0.5$ crossing on each sweep. Widths are calculated for each realization before ensemble averaging. Uncertainties denote one standard error. Each ensemble contains $32$ realizations. Threshold means use the realizations with the corresponding crossing: $32$ in all cases except the truncation $C_\downarrow$ and $\Delta C$, which use $31$. The reduced models use the same $32$ initial conditions, while DNS is independent. Inverted loops satisfy $C_\uparrow<C_\downarrow$.}
\label{tab:transition_results}
\begingroup\small
\setlength{\tabcolsep}{4pt}
\begin{tabular}{@{}lcccc@{}}
\toprule
Model & $C_\uparrow$ & $C_\downarrow$ & $\Delta C$ & Inverted loops\\
\midrule
DNS ($512\times512$) & $0.1210\pm0.0007$ & $0.0777\pm0.0018$ & $0.0433\pm0.0016$ & 0\\
12-mode truncation & $0.1252\pm0.0008$ & $0.1184\pm0.0038$ & $0.0069\pm0.0036$ & 14\\
Closed ROM & $0.1227\pm0.0011$ & $0.0800\pm0.0022$ & $0.0427\pm0.0024$ & 0\\
\bottomrule
\end{tabular}
\endgroup
\end{table*}

\subsection{Particle flux}
\label{sec:remaining}
\begin{figure*}[htbp]
\centering
\includegraphics[width=\textwidth]{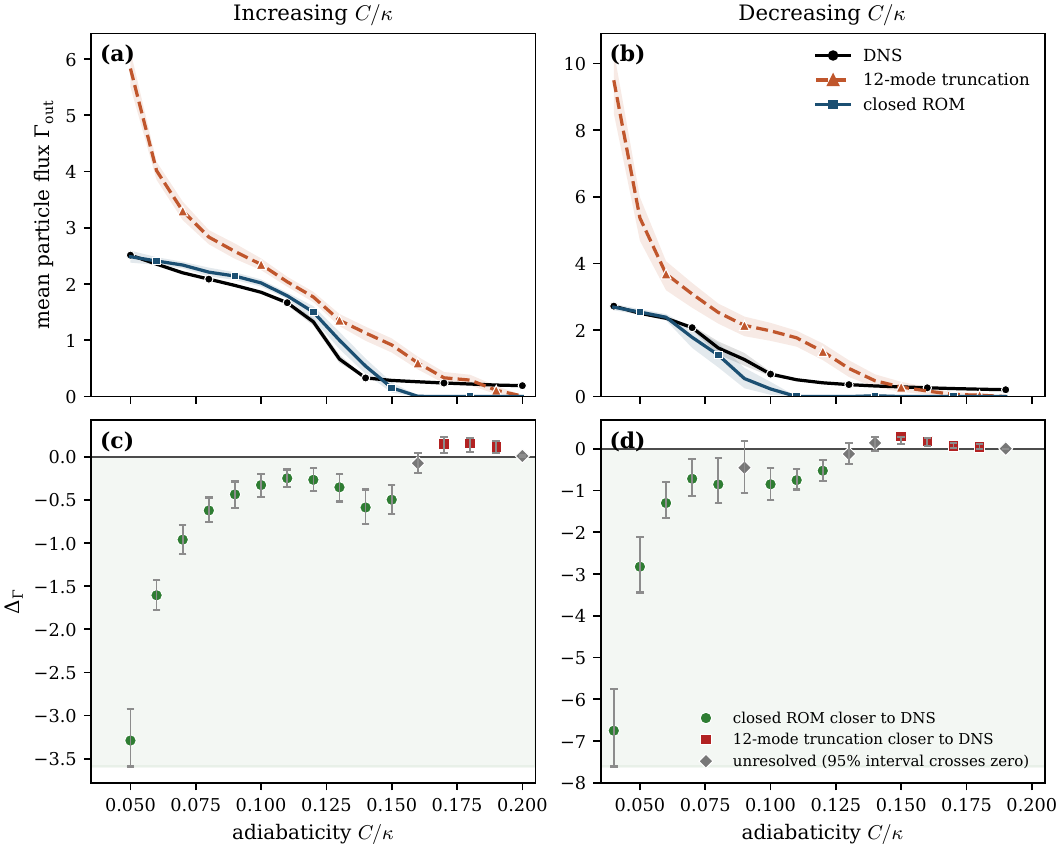}
\caption{Ensemble-averaged radial particle flux on the increasing- and decreasing-$C$ sweeps for DNS, the 12-mode truncation, and the closed ROM, together with the error difference $\Delta_\Gamma$. Across the 32 branch--$C$ conditions, the closed ROM is closer to DNS at 19, the truncation at 7, and the difference is unresolved at 6. Shading in the flux panels and error bars in the $\Delta_\Gamma$ panels show $95\%$ confidence intervals across realizations.}
\label{fig:mean_particle_flux}
\end{figure*}

Particle flux was not used to set the closure parameters and therefore
tests the model beyond hysteresis. We evaluate
\begin{equation}
\Gamma_{\mathrm{out}}
=
-\kappa\left\langle n,\partial_y\phi\right\rangle,
\label{eq:gamma_out}
\end{equation}
where the brackets denote a spatial average of the product. Since $\kappa=1$ throughout, $\Gamma_{\mathrm{out}}$ equals the radial particle flux $\Gamma=-\langle n,\partial_y\phi\rangle$. Figure~\ref{fig:mean_particle_flux} compares the ensemble-mean flux on both sweep branches using the common averaging window of Section~\ref{sec:evaluation_protocol}. The reduced models use the same $32$ initial conditions, while DNS uses $32$ independent realizations. To quantify the change in agreement with DNS, we define
\[
\Delta_\Gamma
=
\left|\overline{\Gamma}_{\rm closure}
-\overline{\Gamma}_{\rm DNS}\right|
-
\left|\overline{\Gamma}_{\rm trunc}
-\overline{\Gamma}_{\rm DNS}\right|,
\]
where the overbar denotes the ensemble mean of the plateau-averaged flux. Negative $\Delta_\Gamma$ indicates a smaller discrepancy for the closure.

Across the 32 branch--\(C\) conditions, the closed ROM is closer to DNS at 19, the truncation at 7, and the difference is statistically unresolved at 6. The closed ROM agrees more closely with DNS at lower $C$. At larger $C$, however, its particle flux falls toward zero while DNS retains finite transport.

\section{Discussion}
\label{sec:discussion}

The results show that feedback from the discarded spectrum can recover much of the hysteresis lost under spectral truncation without increasing the number of retained modes. Particle flux, which was not used to set the parameters, shows that the
improvement is not uniform across the sweep. We first interpret the two
sweep directions, then assess the two working assumptions underlying the closure against DNS, and finally discuss the remaining limitations and their implications for reduced modeling.

\subsection{Physical interpretation of the two sweep directions}
\label{sec:truncation_and_closure}

DWs generate ZFs through nonlinear energy transfer, while ZFs shear the fluctuations and reorganize their spectrum~\cite{dewar2007,pushkarev2013,diamond2005}. The two sweep directions probe different stages of the same turbulence--flow interaction. On the increasing-\(C\) branch, a zonal state forms from a DW-dominated state; on the decreasing-\(C\) branch, DWs regrow about an established zonal state~\cite{numata2007,guillon2025,grander2024}.

As $C$ increases from the turbulent state, strong DW activity supplies the reservoir and sustains the reservoir-dependent ZF damping, $-\nu_{\mathrm{ZF}}\mathcal E\,\widetilde{\mathbf u}_Z$. This feedback opposes ZF growth and therefore acts directly on the formation of the zonal state.

On the decreasing sweep, the system starts in a strongly zonal state
with weak DW activity. The reservoir is then weakly driven and relaxes,
reducing the zonal damping. The zonal field remains strong, however, so
the DW damping, which increases with first-harmonic zonal energy, continues to suppress
perturbations and delays their regrowth.

\subsection{DNS assessment of the closure assumptions}
\label{sec:closure_dns_diagnostics}

\begin{figure*}[htbp]
\centering
\includegraphics[width=\textwidth]{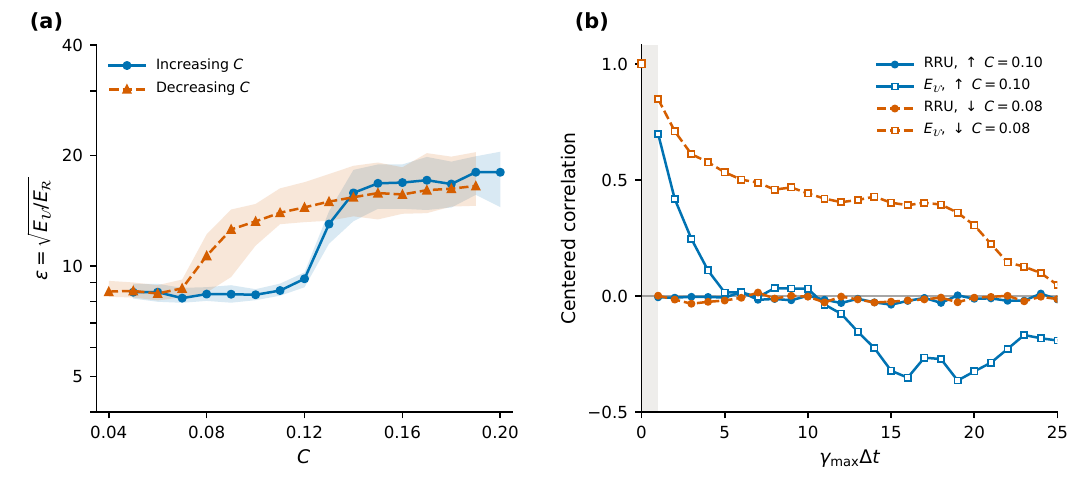}
\caption{(a) Unresolved-to-resolved amplitude ratio $\varepsilon=\sqrt{E_{\mathcal U}/E_{\mathcal R}}$ along the increasing- and decreasing-$C$ sweeps. Curves denote the median across realizations and shading the interquartile range. (b) Centered temporal correlations of the RRU forcing and unresolved energy at $C=0.10$ on the increasing sweep and $C=0.08$ on the decreasing sweep.}
\label{fig:dns_closure_diagnostics}
\end{figure*}

Figure~\ref{fig:dns_closure_diagnostics} assesses the amplitude ordering and short-memory approximation used in the closure. Under the nearest-grid resolved--unresolved partition, the amplitude ratio
$\varepsilon=\sqrt{E_{\mathcal U}/E_{\mathcal R}}$
has a median of $12.41$ and a minimum of $4.78$ across the sampled late-time states. These values support the assumed ordering $\varepsilon\gg1$ for this partition. The energy ratio alone, however, does not establish the relative magnitudes of the feedback terms, which also depend on the interacting wavevectors and phase correlations.

The nearest-grid mapping does not preserve all ROM triads, so the RRU correlation in Fig.~\ref{fig:dns_closure_diagnostics}(b) is an approximate diagnostic of the ROM partition. The temporal correlations are consistent with a faster variation of the RRU forcing than of the unresolved energy. At $C=0.10$ on the increasing branch and $C=0.08$ on the decreasing branch, the forcing correlation is already close to zero at the first nonzero lag, $\gamma_{\max}\Delta t\simeq1$. The corresponding unresolved-energy correlations remain $0.70$ and $0.85$. Because the forcing decay at shorter lags is unresolved, these measurements support a qualitative separation of timescales.

\subsection{Limitations and outlook}
\label{sec:cost_fidelity}

The closure remains phenomenological. Its reservoir source retains only DW--DW excitation, the net relaxation is represented by a constant positive rate, and the resolved feedback is approximated by scalar damping. The exact resolved--unresolved energy balance guides the construction, while the source and damping terms are modeled separately. The reservoir is therefore an effective measure of unresolved activity.

Particle flux was not used to set the closure parameters and therefore provides a separate test beyond hysteresis. The closed ROM agrees more closely with DNS at lower \(C\), while at larger \(C\) its particle flux falls toward zero and DNS retains finite transport. The DW damping is set by the instantaneous first-harmonic zonal energy \(E_{Z,1}\), whereas discarded-mode fluctuations can also repopulate the DW sector. The remaining discrepancies therefore point to feedback missing from this scalar damping model.

More fundamentally, the closure assumes that the discarded-spectrum feedback can be represented in terms of the retained modes and the single reservoir variable \(\mathcal E\). The remaining discrepancies leave open whether this reduced state contains enough information to distinguish the turbulent and zonal-flow-dominated dynamics. Motivated by the history-dependent zonal states and hysteretic dynamics reported in gyrofluid MHW models~\cite{grander2024,grander2026merging}, a natural next step is to reconstruct the corresponding turbulent and zonal-flow-dominated attractors and determine what additional coordinates, if any, are required to describe their dynamics. This tests the state representation itself, rather than introducing additional feedback laws within the existing state.

The MHW system does not contain the equilibrium, geometric and transport physics required for a complete description of the L--H transition. Its role here is instead to provide a minimal setting for testing reduced models of hysteretic turbulence--flow transitions. The results show that a spectral truncation can retain the modes required for the transition while losing feedback from the discarded spectrum that is important for its hysteresis. Representing that feedback through one additional coarse-grained variable recovers much of the lost transition dynamics without enlarging the retained spectral system. Extending this approach to more complete plasma models will require identifying both the reduced state and the discarded-scale feedbacks needed to reproduce transition thresholds and transport.
\section{Conclusion}
\label{sec:conclusion}

We developed a closure for the 12-mode MHW system to recover the hysteresis lost under spectral truncation without adding retained Fourier modes. Guided by the Mori--Zwanzig framework, the closure adds a reservoir that damps the retained zonal modes and a separate DW damping that increases with the first-harmonic zonal energy. The original 12-mode triads remain unchanged.

For the sweeps considered here, the closed ROM reproduces the DNS
hysteresis width, and both transition thresholds lie close to the DNS
values. Particle flux was not used to set the model parameters and therefore provides a separate test beyond hysteresis. The closed ROM recovers the hysteresis and improves particle-flux agreement over much of the sweep; at large \(C\), the finite transport retained by DNS points to additional physics in strongly zonal states.

The remaining discrepancies motivate direct tests of the reservoir source and feedback laws against DNS, together with an assessment of whether additional state variables are needed. The MHW system provides a minimal setting for these tests and for developing reduced descriptions of turbulence--flow feedback relevant to L--H transition physics. Taken together, these results show that a compact closure can recover the hysteretic transition lost under spectral truncation while preserving the original 12-mode triadic structure.

\begin{acknowledgments}
This work was funded by National Research Foundation Singapore (NRF) core funding “Fusion Science for Clean Energy”, under the framework of Singapore Alliance with France for Fusion Energy (SAFE).

\end{acknowledgments}
\section*{Author declarations}

\subsection*{Conflict of Interest}
The authors declare no conflicts of interest.

\subsection*{Author Contributions}
\textbf{Ameir Shaa:} Conceptualization, Data curation, Formal analysis, Investigation, Methodology, Software, Validation, Visualization, Writing – original draft, Writing – review \& editing.
\textbf{Kyungtak Lim:} Conceptualization, Methodology, Resources, Supervision, Writing – review \& editing.
\textbf{Claude Guet:} Conceptualization, Project administration, Resources, Funding acquisition, Supervision, Writing – review \& editing.
\textbf{Xavier Garbet:} Conceptualization, Project administration, Resources, Funding acquisition, Supervision, Writing – review \& editing.

\section*{Data availability}
The data supporting the findings of this study are available from the corresponding author upon reasonable request.
\clearpage
\appendix
\section{Supporting derivations}
\label{app:invariants_couplings}
\label{app:closure_assumptions}

\subsection{Channel notation}
\label{app:couplings}
With $\chi_X$ the indicator of mode set $X$, the ordered coupling, expressed in the transformed coordinates, of inputs from $X$ and $Y$ into the output set $S$ is
\begin{equation}
\begin{aligned}
\mathbf c_{\mathbf k}^{(\mathrm{XY})\mathrm S}
&=\mathbf V^{-1}(\mathbf k)\chi_S(\mathbf k)
\sum_{\mathbf p+\mathbf q=\mathbf k}
\chi_X(\mathbf p)\chi_Y(\mathbf q)\\
&\qquad\times(\mathbf p\times\mathbf q)\phi_{\mathbf p}
\begin{pmatrix}\Omega_{\mathbf q}\\n_{\mathbf q}\end{pmatrix},\\
&\qquad X,Y,S\in\{\mathrm R,\mathrm U\}.
\end{aligned}
\label{eq:cXYZ}
\end{equation}
A channel label lists the two input sets followed by the output set, and a mixed channel contains both input orderings:
$\mathbf c^{\mathrm{RUR}}=\mathbf c^{(\mathrm{RU})\mathrm R}+\mathbf c^{(\mathrm{UR})\mathrm R}$ and
$\mathbf c^{\mathrm{RUU}}=\mathbf c^{(\mathrm{RU})\mathrm U}+\mathbf c^{(\mathrm{UR})\mathrm U}$.

\subsection{Reservoir-source restriction}
\label{app:A2}

Let $\mathbf y$ denote the unresolved response generated from zero by
the RRU forcing,
\[
\dot{\mathbf y}
+
\mathbf M_{\mathcal U}\mathbf y
=
\mathbf c^{\mathrm{RRU}},
\qquad
\mathbf y(0)=0.
\]
The unresolved state can then be decomposed exactly as
\[
\widetilde{\mathbf u}^{\mathcal U}
=
\mathbf y+\mathbf r,
\]
where $\mathbf r$ contains the pre-existing unresolved state and all
contributions not generated by the RRU forcing.

The RRU contribution to the unresolved-energy rate,
$I_{\mathrm{RRU}}\equiv2\operatorname{Re}\langle\widetilde{\mathbf u}^{\mathcal U},\mathbf c^{\mathrm{RRU}}\rangle_{\widetilde{\mathbf G}}$,
therefore splits as
\begin{equation}
I_{\mathrm{RRU}}=J_{\mathrm{RRU}}
+2\operatorname{Re}\langle\mathbf r,\mathbf c^{\mathrm{RRU}}\rangle_{\widetilde{\mathbf G}},
\label{eq:A2_response_remainder}
\end{equation}
where $J_{\mathrm{RRU}}=2\operatorname{Re}\langle\mathbf y,\mathbf c^{\mathrm{RRU}}\rangle_{\widetilde{\mathbf G}}$ is the work of the RRU forcing on its own accumulated response.
The reservoir source is constructed from $J_{\mathrm{RRU}}$; the adopted $(E_{\mathrm{DW}}^{\mathcal R})^2$ law retains its DW--DW contribution.

At fixed zonal amplitude, scaling the DW amplitude by $\delta$ gives
$\mathbf c^{\mathrm{RRU}}=\delta\mathbf F_{ZD}+\delta^2\mathbf F_{DD}$, where $\mathbf F_{ZD}$ and $\mathbf F_{DD}$ are the mixed ZF--DW and DW--DW forcings at unit DW amplitude.
Substituting into this response work orders it as
\[
J_{\mathrm{RRU}}=\delta^2J_{ZD,ZD}+\delta^3\bigl(J_{ZD,DD}+J_{DD,ZD}\bigr)+\delta^4J_{DD,DD},
\]
where $J_{XY}=2\operatorname{Re}\int_0^t\langle\mathbf F_X(t),G_{\mathcal U}(t,s)\mathbf F_Y(s)\rangle_{\widetilde{\mathbf G}}\,ds$ is the work of forcing $\mathbf F_X$ on the propagated response to $\mathbf F_Y$, and $G_{\mathcal U}(t,s)$ propagates the unresolved linear response from $s$ to $t$.
The source $\gamma_{\mathrm{in}}(E_{\mathrm{DW}}^{\mathcal R})^2\propto\delta^4$ of Eq.~\eqref{eq:RRU_source_model} represents $J_{DD,DD}$, the DW--DW contribution to the exact input $I_{\mathrm{RRU}}$ of Eq.~\eqref{eq:A2_response_remainder}.
\section{Numerical reproducibility}
\label{app:benchmark}

\subsection{Numerical details}
\label{app:numerical_conventions}
The DNS viscosity in Eq.~\eqref{eq:mhw} is
\begin{equation}
\nu_{\mathrm{DNS}}(C)=0.017\,\frac{\gamma_{\max}(C)}{k_{y0}(C)^2},
\label{eq:nu_dns}
\end{equation}
following Guillon and G\"urcan~\cite{guillon2025}.
DNS uses a $512\times512$ grid on a square periodic domain of side $32\pi$ with $3/2$ dealiasing and $\kappa=1$, integrated by DOP853 with relative and absolute tolerances $10^{-6}$ and $10^{-12}$ and maximum step $0.1$.
Section~\ref{sec:evaluation_protocol} specifies the sweep, plateau duration, warmup, continuation, ROM integration and crossing rule.

Each ROM realization starts with amplitude $10^{-8}$ in every retained mode except $(0,2)$, which starts with amplitude $10^{-6}$; $\phi_{\mathbf k}$ and $n_{\mathbf k}$ receive independent phases drawn uniformly from $[0,2\pi)$.
Each DNS realization starts from
$\phi_{\mathbf k}=n_{\mathbf k}=10^{-4}\exp[-k^2/(2\sigma_k^2)]\,e^{i\theta_{\mathbf k}}$
with $\sigma_k=0.5$, one uniformly distributed phase $\theta_{\mathbf k}$ shared by both fields, and the spatially uniform mode removed.

DNS stores $201$ samples per plateau and the ROM stores $101$.
In Fig.~\ref{fig:headline}(a), the DNS band is $\bar{\Xi}_K\pm s_{\rm pool}/\sqrt{32}$, where $s_{\rm pool}$ is the root-mean-square deviation of all $32\times201$ full-plateau samples from their pooled mean; the reduced-model bands are standard errors of the realization means.
Negative individual widths are retained in the width averages.
The particle-flux averages use the $49$ samples per plateau at $\gamma_{\max}(C)(t-t_{\mathrm{step}})=102,104,\ldots,198$.
Pointwise $95\%$ intervals for the flux and for $\Delta_\Gamma$ are percentiles of $5000$ whole-realization bootstrap replicates, with DNS resampled independently of the paired reduced-model realizations.

\label{app:seeds}
The reported ROM results use 32 realizations that completed all 32 plateaus with finite $\Xi_K$, selected by ranking the mean absolute deviation of their closed-ROM $\Xi_K$ plateau averages from the DNS mean hysteresis curve; the truncation uses the same initial conditions.
DNS uses an independent ensemble of $32$ realizations.

\subsection{Parameter selection}
\label{app:calibration}
\label{app:observables}
We fix $\gamma_c=1$ and determine the reservoir coupling from the DNS
hysteresis-loop area, with $\eta_{\mathrm{DW}}=0$. We define
\begin{equation}
A_{\mathrm{loop}}
=
\operatorname{trapz}_{\mathcal C_\uparrow}
\left|
\overline{\Xi}_K^\uparrow(C)
-
\mathcal I_\downarrow
[\overline{\Xi}_K^\downarrow](C)
\right|,
\label{eq:benchmark_loop_area}
\end{equation}
where $\operatorname{trapz}_{\mathcal C_\uparrow}$ denotes trapezoidal
integration over the increasing-$C$ grid and $\mathcal I_\downarrow$
interpolates the decreasing-branch mean curve onto the same grid.

The calibration used a $3\times4$ grid in the reservoir-source and
zonal-feedback coefficients. In the $W$-form of the final ROM, these
12 settings correspond to nine distinct effective couplings,
\[
\hwromtwocolumn{%
\begin{aligned}
\Lambda\in{}&\{5.625,\,7.5,\,9.375,\,10,\,11.25,\\
&\quad12.5,\,15,\,15.625,\,18.75\}\times10^{-5}.
\end{aligned}
}{%
\Lambda\in
\{5.625,\,7.5,\,9.375,\,10,\,11.25,\,12.5,\,15,\,15.625,\,18.75\}
\times10^{-5}.
}
\]
The closest agreement with the DNS hysteresis-loop area occurs at
\[
\Lambda=7.5\times10^{-5}.
\]

For the DW damping coefficient, refinement against the DNS transition
thresholds selected
$\eta_{\mathrm{DW}}=3.5\times10^{-3}$ and
$4.0\times10^{-3}$.
We adopt their midpoint,
$\eta_{\mathrm{DW}}=3.75\times10^{-3}$.
Particle flux was not used in parameter selection.

\subsection{DNS diagnostics}
\label{app:dns_orderings}
For Fig.~\ref{fig:dns_closure_diagnostics}, the resolved DNS partition is the set of nearest-grid images of all retained ROM modes, including the eight DW sidebands.
Zonal modes map to $|k_x|=0.5,0.9375$ with $k_y=0$; DW carriers to $k_y=0.375,0.8125$; the sidebands to $k_x=\pm0.5,\pm0.9375$ at those $k_y$; all remaining nonzero modes form $\mathcal U$.

Panel (a) uses $32$ DNS realizations, $32$ plateaus and $11$ late-time frames per plateau, at stored frames $100,110,\ldots,200$, giving $32\times32\times11=11264$ sampled states.
For each realization and plateau, the plotted value is the median over the $11$ sampled frames; the curves and shading show the median and interquartile range across realizations.
The pooled median and minimum quoted in Section~\ref{sec:closure_dns_diagnostics} refer to all $11264$ sampled states.

Panel (b) uses the $101$ consecutive stored frames $100$--$200$ of DNS realization $0$ at $C=0.10$ on the increasing sweep and realization $15$ at $C=0.08$ on the decreasing sweep, with $\gamma_{\max}\Delta t\simeq1$ between adjacent frames.
For a record $X_j$, $j=0,\ldots,N-1$, with $N=101$ and whole-record mean $\bar X$, the lag-$\ell$ correlation is
\[
\rho_X(\ell)=
\frac{(N-\ell)^{-1}\sum_{j=0}^{N-\ell-1}\langle X_j-\bar X,X_{j+\ell}-\bar X\rangle}
{N^{-1}\sum_{j=0}^{N-1}\|X_j-\bar X\|^2}.
\]
For the RRU forcing, $X_j=\mathbf c^{\mathrm{RRU}}$ at frame $j$ with the energy inner product; for the unresolved energy, $X_j=E_{\mathcal U}$ with the scalar product.
\clearpage
\let\reviewbibfont\bibfont
\renewcommand{\bibfont}{\reviewbibfont\setlength{\baselineskip}{9.3pt}}
\bibliography{hw_rom}
\end{document}